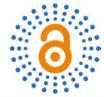

# "Natural Site-Directed Mutagenesis" Might Exist in Eukaryotic Cells

**Gao-De Li**

Chinese Acupuncture Clinic, Liverpool, UK
Email: gaode_li@yahoo.co.uk

## Abstract

**Site-directed mutagenesis refers to a man-made molecular biology method that is used to make genetic alterations in the DNA sequence of a gene of interest. But based on our recently published experimental findings, we propose that "natural site-directed mutagenesis" might exist in eukaryotic cells, which is triggered by harmful agents and co-directed by special transcription hotspots and mutation-contained intranuclear primers.**

## Keywords

**Natural Site-Directed Mutagenesis, Genetic Alteration, Mutation, CAGFs, DNA to DNA Transcription, Drug Resistance, Evolution of Eukaryotes, Antibody Diversity, Somatic Hypermutation**

## 1. Introduction

Site-directed mutagenesis refers to a molecular biology method that is used to make genetic alterations in the DNA sequence of the gene of interest [1] [2]. It is also called oligonucleotide-directed mutagenesis because a mutation-contained oligonucleotide or primer is critical for introducing mutation(s) into the gene of interest. The primer has three functions in the site-directed mutagenesis: first, carrying desired mutations; second, targeting the gene of interest through complementary binding (annealing); and third, serving as a starting point for DNA synthesis. Reasonably, if naturally generated mutation-contained primers are found inside the nucleus of eukaryotic cells, it will be highly possible that a mechanism of "natural site-directed mutagenesis" exists in eukaryotic cells.

Recently we published some novel research findings in which eukaryotic cell, *Plasmodium falciparum*, is found to release certain cell-cycle-associated amplified genomic-DNA fragments (CAGFs) into the nucleoplasm during cell cycle progression, and harmful agent, such as chloroquine, can induce the production of CAGFs. Since CAGFs might belong to single-stranded DNA fragments or molecules that are synthesized through a process of DNA to DNA transcription [3]. It is reasonable to postulate that various single-stranded CAGFs could generate lots of different-sized DNA fragments with variety of DNA sequences through their partial degradation or "programmed splicing or fragmentation". These short single-stranded DNA fragments could be considered as intrinsic or intranuclear primers (IPs, singular form = IP), some of which might serve as ready-to-use "mutation-contained IPs" for certain genes. Therefore, a hypothesis for "natural site-directed mutagenesis" in eukaryotic cells is proposed in this paper.

## 2. The Hypothesis

We hypothesize that "natural site-directed mutagenesis" might exist in eukaryotic cells, which is more efficient than random mutagenesis in the production of genetic alterations in the eukaryotic cells that are constantly exposed to harmful agents. The "natural site-directed mutagenesis" is triggered by harmful agents and co-directed by special transcription hotspots and mutation-contained IPs.

## 3. Mutations in Four Chloroquine Resistance-Associated Genes of *P. falciparum* Could Result from "Natural Site-Directed Mutagenesis"

Till now four chloroquine (CQ) resistance-associated genes, pfmdr1, pfcrt, cg2, and Pfcrmp, have been identified in *P. falciparum* [4]-[7]. There are many mutations in each of these 4 genes: 5 point mutations in pfmdr1 [4], 8 point mutations in pfcrt [5], 12 point mutations and 3 additions or 4 deletions depending on isolates in cg2 [6], and few point mutations and 4 deletions and 2 additions in Pfcrmp [7]. Of the 4 genes, pfcrt has been intensively studied, which has revealed 32 point mutations in pfcrt [8]. These hypergenetic alterations in each of the 4 CQ resistance-associated genes can not be explained by random-mutation theory. Therefore, it is possible that the four mutated genes could result from repeated "natural site-directed mutagenesis". Coincidentally, CQ is found to induce the production of CAGFs in *P. falciparum*. Originally, we think that CAGFs may play an important role in the regulation of gene expression during cell cycle progression [3], but now we speculate that they are also involved in generating mutations in the genome of eukaryotic cells through a mechanism of "natural site-directed mutagenesis" which will be described in more detail below.

"Natural site-directed mutagenesis" in eukaryotic cells is a rare event, and usually results from long-term harmful-agent pressure. Harmful agent, such as CQ, can induce the eukaryotic cells, such a malaria parasites, to produce related CAGFs from which various IPs are generated, and also affects the functions of certain proteins in the cells, the genome is informed through an unknown intracellular communication to produce more mRNA transcripts from which more proteins are translated to compensate for the loss of normal functions of these proteins. Since the newly synthesized non-mutated proteins will not solve the problem, the protein synthesis pressure exerted by the unknown intracellular communication will remain unchanged, forcing the transcription sites to be constantly open for mRNA transcription. These sites can be named as "special transcription hotspots" (STHs, singular form = STH) where single-stranded DNA regions in the transcription bubble are exposed to IPs for a longer time than those in other transcription sites, providing more chances for IPs to be annealed to one of single-stranded DNA regions. The annealed primer could be one single IP or few IPs, and also could be mutation-free IPs or mutation-contained IPs depending on their DNA sequences. If few IPs are annealed, few DNA fragments will be first synthesized, and then joined together by DNA ligase. Through homologous recombination and semiconservative DNA replication, the mutations will be eventually introduced into the newly synthesized genes.

During each process of "natural site-directed mutagenesis", a small population of cells with various mutated genes will be produced. The mutated genes could be harmful, beneficial or neutral to the cells, but under long-time repeated selection (such as CQ pressure), only the cells that contain advantageous mutations will be selected and survive. If the proteins produced by the mutated genes can make cells to adapt easier to the harmful conditions, the STHs will gradually cool down, otherwise the mutagenesis will continue, and more mutations will be accumulated in the same gene. This is why so many mutations are found in each of the four CQ resistance-associated genes.

Whether the introduced genetic alterations are point mutations, or deletions or additions depends on the size of mutation-contained IP, smaller-sized mutation-contained IP might only be able to introduce point mutation, while bigger-sized mutation-contained IP could introduce deletions and additions. Undoubtedly, a bigger gene will have a bigger STH that might provide more chances for a bigger IP annealing, and a smaller gene will have a smaller STH that might provide more chances for smaller IP annealing, which can be used to explain why both deletions and additions are found in the genetic alterations of bigger genes, such as cg2 and Pfcrmp [6] [7], and only point mutations are found in the genetic alterations of smaller genes, such as pfmdr1 and pfcrt [4] [5].

"Natural site-directed mutagenesis" requires the presence of STHs and higher concentration of mutation-contained IPs, *i.e.*, requires the production of more CAGFs. Therefore, the "natural site-directed mutagenesis" is co-directed by STHs and mutation-contained IPs. The mutations generated by "natural site-directed mutagenesis" is very rare in the normal cells that live under favourable conditions because the cells have no STH in the genome and only have low-level CAGFs in the nucleus.

## 4. Somatic Hypermutation Involved in the Generation of Antibody Diversity Might Result from "Natural Site-Directed Mutagenesis"

Somatic hypermutation (SHM) is involved in the generation of antibody diversity by introducing point mutations in the rearranged variable region (V-region) of immunoglobulin (Ig) genes of activated B cells so that the affinity maturation of antibody against a new antigen is achieved. The mutation rate of SHM is one million-fold higher than the spontaneous mutation rate in somatic cells [9]. It is currently believed that activation-induced deaminase (AID) is a key enzyme in SHM, which involves deamination of cytosine to uracil in DNA and generates DNA lesions that are required for the variety of point mutations [10] [11]. However, recent studies suggest that the somatic hypermutation process is similar to natural mutation of evolution [12] [13]. Furthermore, despite many years of studies, the mechanism by which SHM exclusively targets the V-region of Ig genes remains elusive [9].

As a new idea for testing, we think that SHM might result from "natural site-directed mutagenesis" as described above. Presumably, mature B lymphocyte (B cell) carries universal membrane receptors that can non-specifically bind any antigen except self-antigen. The antigen can activate B cell, and may also induce B cell to produce related CAGFs. The activated B cell undergoes proliferation, during which STHs at V-region of Ig genes will be generated, providing opportunity for mutation-contained IP to be annealed to the single-stranded DNA regions in the V-region of Ig genes, triggering synthesis of point mutation-contained DNA strand. If few IPs are annealed, few DNA fragments are first synthesized and then joined together by DNA ligase, through mechanism of homologous recombination and semiconservative DNA replication, the point mutation will be eventually introduced into the V-region of Ig genes, resulting in changes in the membrane receptors of B cell. After many cycles of such "natural site-directed mutagenesis", thousands of thousands of B cells with different new membrane receptors will be produced, but only the B cells carrying receptors with high antigen-binding affinity will survive the antigen selection. Majority of these survived B cells, when specifically bind to the antigen, will differentiate into plasma cells that secrete a large amount of specific antibody, a small number of these B cells will become memory B cells.

There are four theoretical evidences to indirectly support this idea. First, SHM preferentially targets the V-region of Ig genes, which should be thought of as a kind of "site-directed mutagenesis"; second, very high mutation rate in SHM suggests that the mechanism of SHM is very efficient, which is consistent with "natural site-directed mutagenesis"; third, SHM requires transcription of the target genes [14], and "natural site-directed mutagenesis" requires STH; fourth, the mechanism of SHM targeting is associated with certain DNA sequences in and near Ig genes [15], and "natural site-directed mutagenesis" requires IP complementary to certain DNA sequences of the target genes.

## 5. "Natural Site-Directed Mutagenesis" and Evolution of Eukaryotes

Evolution of living organisms requires mutations in the genome and natural selection. The conventional evolutionary theory assumes that the mutations involved in evolution are randomly generated, but since 1988, a number of publications based on studies of bacterial variation challenge the random-mutation theory [16]-[18]. Presumably, both random mutagenesis and non-random mutagenesis are involved in the evolution of living organisms.

Since eukaryotes have more complex genome than bacteria, it is possible that more sophisticated mechanisms of mutagenesis exist in eukaryotes. The mechanism of "natural site-directed mutagenesis" proposed in this paper could be one of efficient ways in the production of mutations in eukaryotes, which might contribute substantially to the evolution of eukaryotes.

## 6. Implications of the "Natural Site-Directed Mutagenesis"

If the hypothesis is proven to be true, it definitely plays an important role in drug resistance development, antibody diversity, and evolution of eukaryotes. At present, it is difficult to obtain a direct evidence to validate the hypothesis, but indirect evidence can be obtained by examining whether small amounts of DNA are synthesized outside the S phase of the *P. falciparum* intraerythrocytic cycle after CQ treatment, the technique used in this examination could be labelled-nucleotide incorporation or electron microscopy. Besides, according to the hypothesis, delivery of more randomly-synthesized short single-stranded DNA fragments into the nucleus will facilitate development of drug resistant line in eukaryotic cells as this delivery will increase the concentration of IPs that are required for the "natural site-directed mutagenesis.

## 7. Conclusion

The hypothesis of "natural site-directed mutagenesis" in eukaryotic cells is proposed in this paper, which is based on an assumption that CAGFs produced during cell cycle progression can generate various mutation-contained intranuclear primers that introduce mutations into the target gene. Theoretically, "natural site-directed mutagenesis" is more efficient than random mutagenesis in the production of genetic alterations, and thus could be one of important functions of eukaryotic cells. Therefore, experimental validation of the hypothesis is necessary.